\documentstyle[11pt,newpasp,twoside,psfig]{article}
\newcommand{\spin}[1]{\,{}_{#1}^{\vphantom{m}}}   
\renewcommand{\l}{{\bf l}}

\newcommand{\da}{d_A}

\newcommand{\cmb}{\Theta}

\newcommand{\vecl}{{\bf l}}
\newcommand{\vecla}{{{\bf l}_1}}

\newcommand{\intl}[1]{\int {d^2 {\bf l}_#1 \over (2\pi)^2}}

\newlength{\tskip}
\newlength{\colwidth}

\newcommand{\beq}{\begin{equation}}
\newcommand{\eeq}{\end{equation}}
\newcommand{\beqa}{\begin{eqnarray}}
\newcommand{\eeqa}{\end{eqnarray}}

\newcommand{\bn}{\hat{\bf n}}
\newcommand{\bm}{\hat{\bf m}}

\newcommand{\rad}{r}    

\newcommand{\len}{\phi}

\begin{document}

\title{Primordial Gravitational Waves and Inflation: CMB and Direct Detection With Space-Based Laser Interferometers}

\author{Asantha Cooray}

\affil{Department of Physics and Astronomy, 4186 Frederick Reines Hall, University of California, Irvine, CA 92697. E-mail: acooray@uci.edu}

\begin{abstract}
The curl-modes of Cosmic Microwave Background (CMB) polarization probe horizon-scale primordial gravitational waves 
related to inflation. A significant source of confusion is expected from a lensing conversion of polarization
related to density perturbations to the curl mode, during the propagation of photons through the large scale structure.
 Either high resolution CMB anisotropy observations or 21 cm fluctuations at redshifts 30 and higher can be used 
to {\it delens} polarization data and to separate gravitational-wave polarization signature from that of cosmic-shear related signal. 
Separations based on proposed lensing reconstruction techniques for reasonable future experiments 
allow the possibility to probe inflationary energy scales down to 10$^{15}$ GeV. 
Beyond CMB polarization, at frequencies between 0.01 Hz to 1 Hz, 
space-based laser interferometers can also be used to probe the inflationary gravitational wave background.
The confusion here is related to the removal of merging neutron star binaries at cosmological distances.
Given the low merger rate and the rapid evolution of the gravitational wave frequency across this
band, reliable removal techniques can be constructed.  We discuss issues related to joint constraints 
that can be placed on the inflationary models based on CMB polarization information and
space-based interferometers such as the Big Bang Observer.
\end{abstract}


\section{CMB: At Present}

The cosmic microwave background (CMB) is now a well known probe of the early universe. 
The temperature fluctuations in the CMB, especially the so-called acoustic peaks in the angular power spectrum of CMB anisotropies, 
capture the physics of primordial photon-baryon fluid undergoing oscillations in the potential 
wells of dark matter (Hu et al. 1997). The associated physics --- involving the evolution of a single
photon-baryon fluid under Compton scattering and gravity --- are both simple and linear, and
many aspects of it have been discussed in the literature since the early 1970s (Peebles \& Yu 1970; Sunyaev \& Zel'dovich 1970).
The gravitational redshift contribution at large angular scales (e.g., Sachs \& Wolfe 1968) and the photon-diffusion damping
at small angular scales (e.g., Silk 1968) complete this description.

By now, the structure of the first few acoustic peaks 
is well studied with NASA's Wilkinson Microwave Anisotropy Probe (WMAP) mission (e.g., Spergel et al. 2003), while 
 in the long term, ESA's Planck surveyor\footnote{http://astro.estec.esa.nl/Planck/}, 
will extend this to a multipole of $\sim$ 2000. Given the improved resolution and better frequency coverage,
a variety of studies related to secondary anisotropies, e.g., Sunyaev-Zel'dovich effect (Sunyaev \& Zel'dovich 1980;
Cooray, Hu \& Tegmark 2000) is expected.
Beyond temperature fluctuations, detections of the polarization anisotropy, related to density or scalar
perturbations, have now been made by DASI (Kovac et al. 2003; Leitch et al. 2004) and
CBI (Readhead et al. 2004; Cartwright et al. 2005),
while the temperature-polarization power spectrum is measured by WMAP (Kogut et al. 2003).

An interesting result from WMAP data is related to the very large ( $> 10^{\circ}$)
angular scale polarization signal, which probes the local universe and associated astrophysics 
instead of physics at the last scattering surface probed with polarization at
tens of arcminute scales. This local universe contribution arises when the universe reionizes again and
the temperature quadrupole begins to rescatter to produce a new contribution
to the polarization anisotropy at angular scales corresponding to the horizon at the new scattering surface 
(Zaldarriaga 1997). The large-scale excess signal 
measured in the WMAP temperature-polarization cross-correlation
 power spectrum is interpreted as rescattering 
with an  optical depth to electron scattering of $0.17 \pm 0.04$ 
(Kogut et al. 2003).

Such an optical depth suggests early reionization; for example, if the universe reionized completely and instantaneously at
some redshift, the measured optical depth suggests a reionization redshift of $\sim$ 17 ($\pm 5$). 
Such a high optical depth, in the presence of a $>$ 1\% neutral Hydrogen fraction from z $\sim$ 6 Sloan quasars (Fan et al. 2002), suggests a complex reionization history. A complex, and patchy, 
reionization, however, is expected if the reionization is dominated by the
UV light from first luminous objects, though to explain the high level of the WMAP's optical depth requires
high star formation efficiency at redshifts of order 15 
with a possibility for a population of metal-free massive Pop III stars (e.g., Cen 2003).

In general, one can only extract limited information related to the reionization history 
from  polarization peak at tens of degrees or more angular scales (e.g., Hu \& Holder 2003).
One reason for this is the large cosmic variance associated with 
measurements at multipoles of $\sim$ 10. To extract detailed information of the reionization process,
one can move to temperature fluctuations generated during the reionization era at angular scales
corresponding to few arcminutes and below. At these small angular scales, potentially interesting contributions from
high redshifts include a scattering contribution related to moving electrons in ionized patches and
fluctuations in the moving electron population (e.g., Santos et al. 2003),
and a thermal Sunyaev-Zel'dovich effect
related to the first supernovae bubbles (Oh et al. 2003). The same redshift ranges are expected to be
illuminated in the near-infrared band between 1 to 3 $\mu$m in the form of spatial fluctuations to the cosmic
IR background (e.g., Santos et al. 2002; Salvaterra et al. 2003;
Cooray et al. 2004). Potentially interesting studies to extract more information related 
to the reionization process include brightness temperature fluctuations of the redshifted
21 cm line emission by neutral HI prior to and during Hydrogen (e.g., Zaldarriaga et al. 2004),
and cross-correlation studies between small angular scale
effects in CMB and maps at other wavelengths (Cooray 2004a; Cooray 2004b; Salvaterra et al. 2005). 
 Upcoming experiments such as the South Pole Telescope and the
Atacama Cosmology Telescope, combined with near-IR and 21 cm maps, 
 will provide some of the first possibilities in this direction.

\begin{figure}[t]
\centerline{\psfig{file=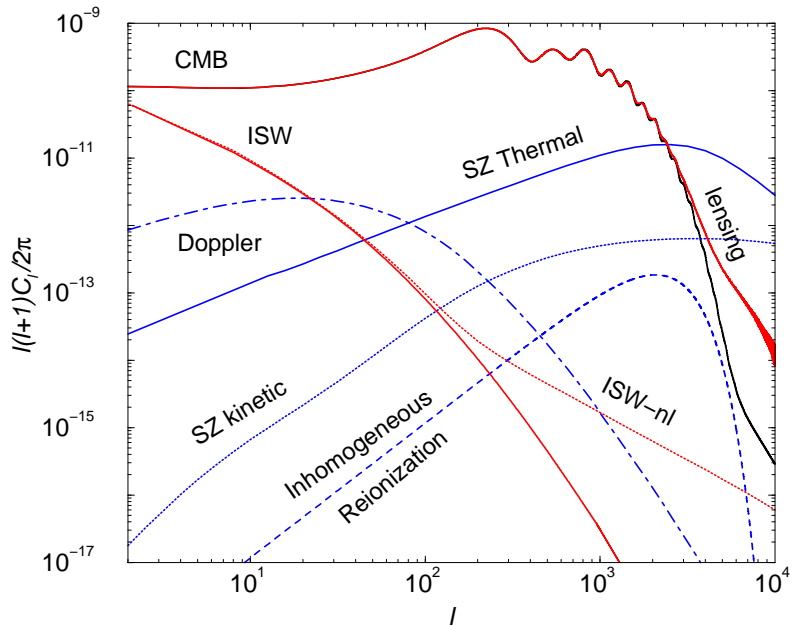,height=20pc,angle=-90}}
\caption{Power spectrum for the temperature 
anisotropies in the fiducial $\Lambda$CDM model
with $\tau=0.1$. In the case of temperature, the curves show the
local universe contributions to CMB due to
gravity (ISW and lensing) and scattering (Doppler, SZ effects,
patchy reionization). We refer the reader to Cooray, Baumann \& Sigurdson (2004) for a recent
review on large scale structure contributions to temperature anisotropies.}
\label{fig:sec}
\end{figure}

\begin{figure}[!h]
\centerline{\psfig{file=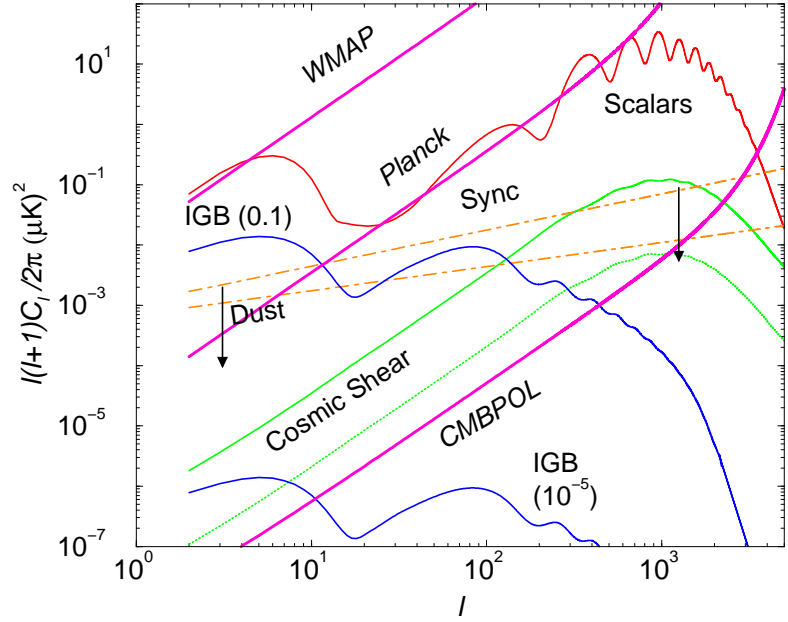,height=20pc,angle=-90}}
\caption{CMB polarization anisotropies in the gradient (E) mode due to scalars and curl (B) modes due to
the inflationary gravitational wave background (curve labeled 'IGB', with the number in parenthesis showing the normalization
in terms of the tensor-to-scalar ratio, which in terms of the energy scale of inflation is
$2 \times 10^{16}$ GeV (0.1), top curve, and $2 \times 10^{15}$ GeV (10$^{-5}$), bottom curve). 
Note the effect due to
reionization where rescattering produces new large angular scale anisotropies. 
For optical depths at the level of 0.1, the peak of the power related to IGW curl modes is at tens of degree angular scales and not
at the degree scale ($l \sim 100$) corresponding to the bump at recombination.
Note that this enhancement, amounting to over two magnitudes in power at $l \sim$ few 
is significant and aid potential detectability of the IGWs via polarization.
The IGW detection is confused with cosmic shear conversion of  a fractional E-mode $\rightarrow$ B-mode 
by the intervening large scale structure.  The residual noise curve related to cosmic shear is after doing lens-cleaning in polarization maps
with an arcminute scale CMB experiment with instrumental polarization sensitivity 
at the level of 1 $\mu$K $\sqrt{\rm sec}$. For reference, we show an estimated level of contamination from
dust and synchrotron emission (at 150 GHz);  this noise level can be improved with foreground cleaning
in multifrequency data with sensitivity better than the foreground noise contributions. 
The curve labeled Planck is the noise curve related to upcoming Planck (HFI) polarization observations as
a function of each multipole.
}
\label{fig:cl}
\end{figure}

\section{Inflation: CMB and Space-Based Laser Interferometers}

We first summarize basic aspects related to inflation under the assumption of a single scalar field model.
Our derivations, however, are
applicable to multi-field theories with few alterations (Lidsey et al. 1997).   In general,
constraints on inflation are obtained by comparing theoretically calculated scalar  and tensor metric --- gravitational waves --- 
perturbations to direct (in the case of space-based detectors) and indirect (such as  CMB)
measurements of these perturbations.  

The tensor and scalar perturbations are determined by the evolution of the scalar field, $\phi$, (the 'inflaton' field) and the scale factor, $a$, with time. 
The equation of motion for a scalar field with a self-interaction potential, $V(\phi)$, is given by, 
\begin{equation}
\ddot{\phi} + 3 H \dot{\phi}+  V^{\prime} = 0,
\end{equation}
where $H$ is the Hubble parameter and the prime refers to differentiation with respect to the field, $\phi$.  The evolution of the 
Hubble parameter is determined by the energy density, $\rho$, which we suppose is dominated by the energy density associated with the inflaton field, 
\begin{equation}
H^2 = \frac{8 \pi}{3 M_{pl}^2} \rho = \frac{8 \pi}{3 M_{pl}^2} \left(\frac{1}{2} \dot{\phi}^2 + V(\phi)\right) \,,
\end{equation}
where $M_{pl} \equiv G^{-1/2} = 1.2 \times 10^{19} \ \mathrm{GeV}$. 
The initial curvature redshifts 
away by the time the scales corresponding to the current Hubble horizon left the inflationary horizon.
In general, an exact
analytic solution for the spectrum of scalar and tensor perturbations does not exist.  In order to 
facilitate analytic solutions, one generally resorts to the 'slow-roll' expansion, 
where the equations of motion are expanded in terms of the inflaton potential and its derivatives 
(Liddle, Parsons \& Barrow 1994).  
Motivation for this expansion can be understood by noting that the equation of motion for the inflaton field is formally equivalent 
to the equation of motion of a mass moving under a potential $V$ with damping due to the cosmological expansion.  The slow roll expansion supposes that the 
motion of the field reaches a 'terminal velocity' in which the potential energy dominates over the kinetic energy and the acceleration of the field is 
negligible.  In order to quantify these assumptions, the slow-roll parameters are used 
\begin{eqnarray}
\epsilon \equiv \frac{M_{pl}^2}{16 \pi} \left(\frac{V^{\prime}}{V}\right)^2 \, , \quad {\rm and} \quad 
\eta \equiv \frac{M_{pl}^2}{8 \pi} \frac{V^{\prime\prime}}{V}. 
\end{eqnarray}
Under  the assumption that $\epsilon, |\eta| \ll 1$, the approximate 'slow-roll' evolution equations are 
\begin{eqnarray}
H^2 \approx \frac{8 \pi}{3 M_{pl}^2} V\, , \quad {\rm with} \quad
3 H \dot{\phi} + V^{\prime} \approx 0 \, .
\end{eqnarray}
On a given length scale,  scalar  perturbations start as quantum fluctuations in otherwise classical fields well within the horizon. Here, flat space-time quantum 
field theory is appropriate.   As the scale factor quasi-exponentially
 increases, the length scale is stretched beyond the inflationary horizon. The quantum 
fluctuations are then amplified and  become classical perturbations to the metric (Guth \& Pi 1982).  
In order to study scalar and tensor perturbations, we consider their power spectra.  
It can be shown that within the slow roll approximation the amplitude of 
power spectra can be written as (Stewart \& Lyth 1993, Lidsey et al. 1997), 
\begin{eqnarray}
P_{S}(k) &=& \frac{1}{\pi} \frac{H^2}{M_{\mathrm{pl}}^2 \epsilon} \approx \frac{128 \pi}{3 M_{\mathrm{pl}}^6}\frac{V^3}{V^{\prime 2}},\\
P_{T}(k) &=&  \frac{16}{\pi M_{\mathrm{pl}}^2}H^2\approx \frac{128}{3}\frac{V}{ M_{\mathrm{pl}}^4}.
\end{eqnarray}
where $S$ and $T$ refer to scalar and tensor perturbations respectively. 
The right hand sides of the above equations are to be evaluated when a given 
comoving length scale crosses the inflationary horizon with $k_0 = aH$.  
In order to quantify how these power spectra change with scale, 
one considers their power-law spectral indices (Liddle and Lyth 1992),
 \begin{eqnarray}
n_S (k) &\equiv& 1- \frac{d \ln P_S}{d \ln k} \approx 1-6\epsilon + 2 \eta, \\
n_T(k) &\equiv& \frac{d \ln P_T}{d \ln k} \approx -2\epsilon.
\end{eqnarray}
If the potential is flat enough, higher order spectral parameters, which are higher order in the slow roll parametersm, can be ignored (Stewart \& Lyth 1993).   
As long as the slow roll parameters can be considered constant, we are able to write the power spectra as power-laws with spectral indices given  by the above expressions. It can then be shown that, 
\begin{equation}
\Delta \epsilon \approx \Delta \eta \approx \mathcal{O}\mathrm(\epsilon^2) \Delta \ln \mathit k.
\end{equation}
Currently, CMB data probe up to $l_{max} \sim 1000$ corresponding to $\Delta \ln k \sim 1$.  Current observations constrain $\epsilon \leq 0.1$ at scales corresponding to the current Hubble horizon.  Therefore, although CMB observations do cover a few orders of magnitude in wavenumber, to within second order in the slow roll parameters, they can be considered constant over the range probed by the CMB.  This implies that the perturbation power spectra will appear as approximate power laws, expanded around some 'pivot' wavenumber, $k_0$ (Leach et al. 2002), 
\begin{eqnarray}
P_S(k) \approx P_S(k_0) \left(\frac{k}{k_0}\right)^{1-n_S} \, , \quad {\rm and} \quad 
P_T(k) \approx P_T(k_0) \left(\frac{k}{k_0}\right)^{n_T} \, .
\end{eqnarray}
In the case of WMAP these parameters are evaluated at $k_0 = 0.002 \mathrm{Mpc}^{-1}$ (Spergel et al. 2003; Peiris et al. 2003).  
Using the expressions for the amplitudes and spectral indices in terms of the slow roll parameters, we are able to solve for the potential and its first two derivatives and thereby use observations to constrain the form that this potential can take.  
We have, 
\begin{eqnarray}
V &\approx& \frac{3 P_T}{128} M_{pl}^4,\\
|V^{\prime}| &\approx& \sqrt{- 8 \pi n_T} \frac{3}{128}P_T M_{pl}^3,  \\
V^{\prime \prime} &\approx& \frac{3 \pi}{32} P_T(n_S-1-3 n_T) M_{pl}^2,\\
\frac{P_T}{P_S} &\approx& - 16 n_T,
\end{eqnarray}
and all of the spectral parameters are to be evaluated at the pivot wavenumber, $k_0$.  The last expression is known as the consistency relation and is true only in single-field inflationary model. The expression relaxes to an inequality in the case of multi-field models.  It is a consequence of the fact that both scalar and tensor spectra are determined from the dynamics of a single scalar field, so they are not independent (Lidsey et al. 1997, 
Liddle \& Lyth 2000).  

\subsection{CMB}

Though acoustic oscillations in the temperature anisotropies of CMB suggest an inflationary origin for
primordial perturbations, such as in the WMAP data (Peiris et al. 2003), it has been argued for a while
 that the smoking-gun signature for inflation would be the detection of a stochastic
background of gravitational waves (e.g., Kamionkowski \& Kosowsky 1999).  
Since CMB probes horizon to super-horizon scales, inflation is the only known mechanism to casually
produce a stochastic background of gravitational waves at such large wavelengths (e.g., Starobinsky 1979).
Note that the amplitude
of the gravitational wave background is highly unknown.
Since the amplitude of these inflationary gravitational waves is proportional to ${\cal V}$,
the value of the
inflaton potential $V(\phi)$ during inflation, the amplitude of gravitational-wave
induced B-mode polarization anisotropies directly constrains the
energy scale of inflation ${\cal V}^{1/4}$.
Relative to COBE  normalization, the
relation between the energy scale and the ratio of tensor-to-scalar
fluctuations is ${\cal V}^{1/4} = 3.0 \times 10^{-3} (T/S)^{1/4}M_{pl}$ 
(Turner \& White 1996), 
The current limit from CMB anisotropy data, based on large angular-scale CMB temperature
fluctuations, suggests an upper limit of order $2 \times 10^{16}$ GeV (Melchiorri \& Odman 2003).

Instead of temperature fluctuations (where dominant scalar perturbations confuse the 
detection  of tensor contribution), the gravitational-waves can be 
studied through their distinct signature in the CMB polarization in the form of a contribution to the curl, 
or magnetic-like, component (Kamionkowski et al. 1997; Seljak \& Zaldarriaga 1997).
There is no contribution from the dominant scalar, or density-perturbation, contribution to these curl modes in linear theory.
Secondary effects related to density perturbations, however, do
produce signals in the curl-mode and confuse the detection of inflationary gravitational-wave signal.

Among these, the few percent conversion of E to B from cosmic shear modification to the
polarization pattern by the intervening large scale structure is important (Zaldarriaga \& Seljak 1998). These
lensing-induced curl modes introduce a noise from which gravitational waves must be distinguished in the CMB polarization.

Gravitational lensing of the CMB photons by the mass fluctuations in the 
large-scale structure is now well understood (e.g., Hu 2000; Hu \& Cooray 2001). 
The lensing effect can be described through a remapping process involving angular deflections
resulting along the photon path:
\begin{eqnarray}
\tilde \cmb(\bn) & = &  \cmb(\bn + \nabla\len) \nonumber\\
        & = &
\cmb(\bn) + \nabla_i \len(\bn) \nabla^i \cmb(\bn) + {1 \over 2} \nabla_i \len(\bn) \nabla_j \len(\bn)
\nabla^{i}\nabla^{j} \cmb(\bn)
+ \ldots \, ,
\end{eqnarray}
where the deflection angle on the sky is given by
the angular gradient of the lensing potential,
$\alpha(\bn) = \nabla \phi(\bn)$, which itself is a
projection of the gravitational potential, $\Phi$ (see e.g. Kaiser 1992),
\begin{eqnarray}
\phi(\bm)&=&- 2 \int_0^{\rad_0} d\rad
\frac{\da(\rad_0-\rad)}{\da(\rad)\da(\rad_0)}
\Phi (\rad,\hat{{\bf m}}\rad ) \,.
\label{eqn:lenspotential}
\end{eqnarray}
The quantities here are the conformal distance or lookback time, from the observer ($\rad$)
and the analogous angular diameter distance $d_A$.
The lensing potential in equation~(16)
can be related to the well known convergence generally
 encountered in conventional lensing
studies involving galaxy shear (Kaiser 1992)
\begin{eqnarray}
\kappa(\bm) & =& {1 \over 2} \nabla^2 \phi(\bm) \nonumber \\
& = &-\int_0^{\rad_0} d\rad \frac{\da(\rad)\da(\rad_0-\rad)}{\da(\rad_0)}\nabla_{\perp}^2 \Phi (\rad ,\
\hat{{\bf m}}\rad) \, , \nonumber \\
\end{eqnarray}
where note that the 2D Laplacian operating on
$\Phi$ is a spatial and not an angular Laplacian.
Expanding the lensing potential to Fourier moments,
\begin{equation}
\phi(\bn) = \int \frac{d^2\vecl}{(2\pi)^2} \phi(\vecl)
{\rm e}^{i \vecl \cdot \bn}  \, ,
\end{equation}
we can write the usually familiar quantities of convergence and shear components of
weak lensing as 
\begin{eqnarray}
\kappa(\bn) &=& -\frac{1}{2}\int \frac{d^2\vecl}{(2\pi)^2}
l^2 \phi(\vecl) {\rm e}^{i\vecl \cdot \bn} \nonumber \\
\gamma_1(\bn) \pm i\gamma_2(\bn) &=& -\frac{1}{2}\int \frac{d^2\vecl}{(2\pi)^2}
l^2 \phi(\vecl) {\rm e}^{\pm i 2 (\varphi_l-\varphi)}{\rm e}^{i\vecl \cdot \bn} \, .
\label{eqn:kappa}
\end{eqnarray}
Taking  the Fourier transform, as appropriate for a flat-sky, we write
\begin{eqnarray}
\tilde \cmb(\vecla)
&=& \int d \bn\, \tilde \cmb(\bn) e^{-i \vecla \cdot \bn} \nonumber\\
&=& \cmb(\vecla) - \intl{1'} \cmb(\vecla') L(\vecla,\vecla')\,,
\label{eqn:thetal}
\end{eqnarray}
where
\begin{eqnarray}
\label{eqn:lfactor}
L(\vecla,\vecla') &=& \len(\vecla-\vecla') \, (\vecla - \vecla') \cdot \vecla'
+{1 \over 2} \intl{1''} \len(\vecla'') \\ &&\quad
\times \len^*(\vecla'' + \vecla' - \vecla) \, (\vecla'' \cdot \vecla')
               (\vecla'' + \vecla' - \vecla)\cdot
                             \vecla' \,.  \nonumber
\end{eqnarray}

The lensed power spectrum, according to the present formulation, is discussed in  Hu (2000)  and we can write
\begin{eqnarray}
\tilde C_l^\cmb &=& \left[ 1 - \intl{1}
C^{\phi\phi}_{l_1} \left(\vecl_1 \cdot \vecl\right)^2 \right]   \,
                                C_l^\cmb
        + \intl{1} C_{| \vecl - \vecl_1|}^\cmb C^{\phi\phi}_{l_1}
                [(\vecl - \vecl_1)\cdot \vecl_1]^2  \, . \nonumber \\
\label{eqn:ttflat}
\end{eqnarray}

Since the lensing effect involves the angular gradient of CMB photons and leaves the surface brightness unaffected, its  signatures are at 
the second order in temperature. Effectively, lensing smoothes the acoustic peak structure at large angular scales and
moves photons to small scales.  Note that the approach above treats the deflection angle
as a perturbative parameter. Such an approximation breaks down at angular scales corresponding to roughly the rms
deflection angle of a CMB photon ($\sim$ a few arcminutes). While the approach can be extended to higher order
(e.g., Cooray 2004c), the numerical calculation becomes increasingly time consuming. The development of
numerically fast, but still accurate, approaches (e.g., Challinor \& Lewis 2005)
will aid in analysis data from future high resolution CMB experiments. To put rough estimates on the
expected changes due to lensing, the mean gradient is of order
$\sim$ 15 $\mu$K arcmin$^{-1}$ and with deflection angles or order 0.5 arcmin or so from massive clusters, the lensing effect results in temperature fluctuations of order $\sim$ 5 to 10 $\mu$K. 
In Fig.~1, we show the effect of lensing on temperature fluctuations where power is
transferred from acoustic peaks to the damping tail of anisotropy spectrum.

\begin{figure}[!ht]
\centerline{\psfig{file=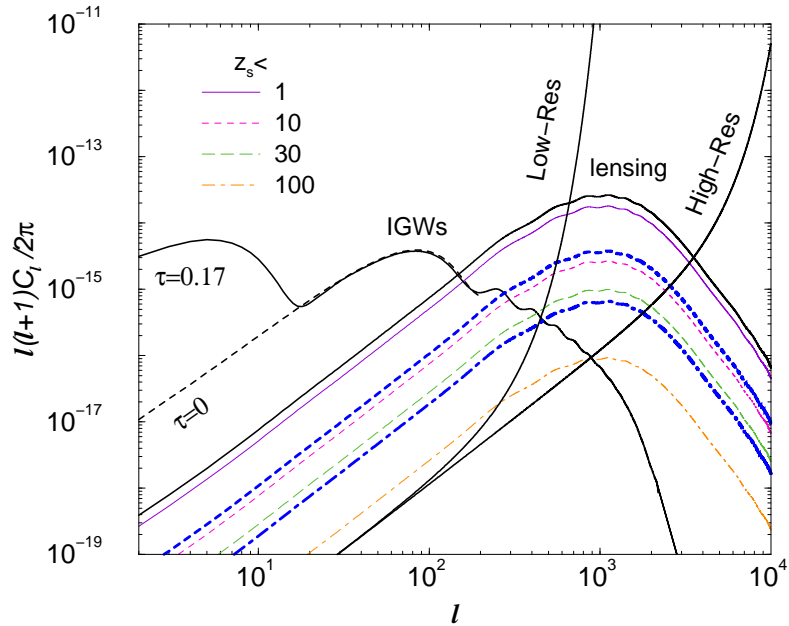,height=20pc,angle=-90}}
\caption{CMB B-mode polarization. The curve labeled `IGWs'
is the IGW contribution with a tensor-to-scalar ratio of 0.1 with (solid line; $\tau=0.17$) and without (dashed line) 
reionization. The curve labeled `lensing' is the total lensing confusion to B-modes. 
Thin lines show the residual B-mode lensing contamination for removal of with lensing out to $z_{s}$ while
thick lines show estimates of the residual confusion from CMB experiments alone using quadratic estimators with an 
ideal noise-free experiment  (dashed line), and likelihood methods using a high resolution/sensitivity experiment (dot-dashed). The noise curve of latter this experiment, 
with 2 arcminute beams and a pixel noise of 0.25 $\mu$K-arcminute, is the curve labeled `High-Res'.  Bias-limited lensing information to  $z_{s} > 10$
improves upon the limit to the IGW amplitude based on quadratic statistics, and the likelihood level can be reached with $z_{s} \sim 30$.  
If $z_{s} \sim 100$ an additional order-of-magnitude in the IGW amplitude could be probed.  The curve labeled `Low-Res' is the noise curve for a 
lower resolution CMB polarization experiment with 30 arcminute beams sufficiently sensitive to detect IGWs when paired with a cosmic 21-cm 
lensing reconstruction. See, Sigurdson \& Cooray (2005) for details.
}
\label{spectra}
\end{figure}

To calculate the lensing contribution to the polarization, we follow the notation in Hu (2000).
Introducing,
\begin{eqnarray}
{}_{\pm} \tilde X(\bn) & = &  {}_{\pm}X(\bn + \nabla\len) \\
\label{eqn:ebl}
        & \approx &
{}_{\pm} X(\bn) + \nabla_i \len(\bn) \nabla^i {}_{\pm} X(\bn) + {1 \over 2} \nabla_i \len(\bn) \nabla_j \len(\bn) \nabla^{i}\nabla^{j}
{}_{\pm}  X(\bn) \,,
\nonumber
\end{eqnarray}
where $\spin{\pm} X = Q\pm i U$ represent combinations of the Stokes parameters. The lensing effect is again to move the photon
propagation directions on the sky.   In Fourier space, assuming flat-sky, we can consider the E- and B-mode decomposition introduced in Kamionkowski et al. (1997; Seljak \& Zaldarriaga 1997) 
such that
${}_\pm X(\l) = E(\l)\pm i B(\l)$ and
\begin{eqnarray}
{}_\pm \tilde X(\l) &=& {}_\pm X(\l)
-
\intl{1}
{}_\pm X(\l_1)
 e^{\pm 2i (\varphi_{\vecl_1}- \varphi_{\vecl})} L(\l,\l_1) \,.
\end{eqnarray}
Since primordial $B(\vecl)$, due to the gravitational wave contribution, is small, we can make the 
useful approximation that
\begin{eqnarray}
\tilde E(\vecl) &=& E(\vecl) - \intl{1} \left[E(\vecl_1) \cos 2 (\varphi_{\vecl_1} -\varphi_\vecl)\right] \phi(\vecl-\vecl_1) (\vecl-\vecl_1) \cdot \vecl_1 \nonumber \\
\tilde B(\vecl) &=& - \intl{1} \left[E(\vecl_1) \sin 2 (\varphi_{\vecl_1} -\varphi_\vecl)\right] \phi(\vecl-\vecl_1) (\vecl-\vecl_1) \cdot \vecl_1 \, ,
\end{eqnarray}
Under this approximation, the lensed polarization power spectra can be
expressed in terms of $C_l^{\phi\phi}$ and, in the case of B-modes, we can write
\begin{eqnarray}
\tilde C_l^{BB} &=& C_l^{BB, IGW} +
        \frac{1}{2} \intl{1} C_{|\vecl - \vecl_1|}^{\phi\phi}
[(\vecl - \vecl_1)\cdot \vecl_1]^2
C^{EE}_{l_1} [1 - \cos(4\varphi_{\vecl_1})]  \, ,\nonumber \\
\end{eqnarray}
where $C_l^{BB,IGW}$ is the contribution due to IGWs and the contribution in the second
line is the confusion that must be removed. To achieve this separation, one must 
obtain information on the integrated potential out to last scattering surface.
One can effectively extract information related to the lensing potential, and the
integrated dark matter density field responsible for the lensing effect, via 
 quadratic statistics applied to CMB temperature and polarization 
maps (Hu 2001a, Hu 2001b; Hu \& Okamoto 2003; Cooray \& Kesden 2003; Kesden et al 2003) 
or likelihood-based techniques (Hirata \& Seljak 2003). 

In figure~2, we show the contribution from dominant scalar modes to the polarization in the gradient
 component and from gravitational-waves to the curl polarization. 
Unlike the B modes generated by tensor perturbations (the IGWs), 
$C_l^{EE}$ is dominated by larger amplitude scalar perturbations.  
The expected few-percent
 conversion of E modes creates a large signal in the B-mode power spectrum (Zaldarriaga \& Seljak
1998). For tensor-to-scalar ratios $T/S$ below $2.6 
\times 10^{-4}$ or 
${\cal V}^{1/4}$ below $4.6 \times 10^{15}$~GeV the IGW signal is completely confused by the lensing contaminant (Lewis et al. 2002).
To bypass this limit one must separate the lensing induced B-modes from those due to IGWs (Kesden et al. 2002; Knox \& Song 2002;
Seljak \& Hirata 2004).
Clearly, the lensing confusion could be exactly removed if one knew the three-dimensional distribution of mass out 
to the CMB last-scattering surface.
However,  knowledge of the projected quantity $\phi({\hat{\bf n}})$ is sufficient. One way to estimate 
$\phi({\hat{\bf n}})$ is by using quadratic estimators or maximum likelihood methods to statistically infer the deflection-angle field given temperature and polarization anisotropies (Hu 2001b; Hu \& Okamoto 2002). These method allow one to detection down $10^{15}$ GeV (Kesden et al. 2002; Knox \& Song 2002). In the case of likelihood methods, one does not have an obvious ultimate limit, however, the
residual lensing noise is still above the detector noise even for highest sensitive polarization maps.
While this is encouraging, it is useful to consider alternative techniques to reduce the confusion.
A useful way to estimate $\phi({\hat{\bf n}})$, but out to a high-redshift, 
is by observing the weak-lensing distortions of the shapes of objects of a known average shape at redshift $z_s$.
As the CMB source redshift is at $1100$, observations of the weak lensing of galaxies, out to a redshift of $\sim 1$--$2$, 
can not be used to effectively delens CMB maps because a 
large fraction of the lensing contamination (55\% at $l=1000$) comes from structure at $z > 3$.
A possibility exists in the form anisotropies in the
21 cm brightness temperature fluctuations during the era and prior to reionization (Sigurdson \&
Cooray 2005).

For example, using a quadratic reconstruction of the deflection field 
of 21 cm anisotropies out to a source redshift of 30, in conjunction with a low-resolution
but high signal-to-noise CMB polarization 
experiment, could detect $T/S > 2.5 \times 10^{-5}$ or ${\cal V}^{1/4} > 2.6 \times 10^{15}$~GeV if 21 cm fluctuations measurements are limited 
to $l \sim 5000$. If the
resolution in 21 cm observations is improved so that one can measure 21 cm brightness
temperature anisotropy spectrum out to $l \sim 10^5$, one can improve down to
$T/S > 1.0 \times 10^{-6}$ or ${\cal V}^{1/4} > 1.1 \times 10^{15}$~GeV.
This limit is comparable to using CMB lensing information alone from high sensitive
maps at 2 arcminute resolution (Seljak \& Hirata 2004).  To achieve such a resolution
one requires roughly a space mission with an aperture of diameter $\sim 6 (100 \; {\rm GHz}/\nu)$ meters.
In the case of CMB lens-cleaning with information from 
21 cm observations, as the IGW signal is at multipoles of 100 (in the
case of recombination bump) and/or 20 (in the case of reionization  with optical
depths around 0.1), the CMB data need not be high resolution.
Since lensing information is not  extracted from CMB 
observations, the optimal observing strategy is to integrate over a few square degree patch of 
the sky as proposed in Kamionkowski \& Kosowsky (1998).  
Thus, it is conceivable that the 
Inflationary Probe of NASA's Beyond Einstein Program can be designed in combination with a cosmic 21-cm radiation experiment.  
Such a cosmic 21-cm radiation experiment could be extremely exciting for reasons involving 21 cm fluctuations alone where 
one probes the small scale structure of the
Universe at high redshifts.
For a bias-limited reconstruction out to a $z_s \sim 100$, the limit on the tensor-to-scalar ratio is $7.0 \times  10^{-8}$ or  ${\cal V}^{1/4} > 6.0 \times 10^{14}$ GeV.  For $z_s \sim 200$, the maximum redshift where 21-cm fluctuations are expected to be nonzero, one could probe down to ${\cal V}^{1/4} > 3 \times 10^{14}$ GeV.

\subsection{Space-based Laser Interferometers}

Beyond CMB polarization data, there is long-term interest, as part of NASA's Beyond Einstein Program, to
launch a space-based laser interferometer (Big Bang Observer) to directly detect the presence of relic stochastic gravitational
background today from inflation. The CMB detection and the detection  of the laser interferometers provide
unique opportunities to study properties related to inflationary physics. To understand requirements for the
direction detection, we summarize some of the key aspects related to the stochastic background.

The evolution of tensor perturbations, for scales inside the horizon subsequent to inflation, is governed by 
the massless Klein-Gordon equation
\begin{equation}
\frac{d^2h_k}{d \tau^2}+ 2 \frac{1}{a} \frac{da}{d \tau} \frac{dh_k}{d \tau} + k^2 h_k = 0
\end{equation}
subject to the boundary conditions, $h_k(0) = P_T(k)^{1/2}$ and $\dot{h_k}(0) = 0$.  $\tau$ is conformal time, which is related to cosmic time by the differential relation, $dt/d\tau = a$.
We define $g_k(\tau) \equiv h_k(\tau) a(\tau)$.  With this definition, we are able to rewrite the above equation,
\begin{equation}
\frac{d^2 g_k}{d \tau^2} +\left(k^2 - \frac{1}{a} \frac{d^2 a}{d \tau^2} \right) g_k = 0
\end{equation}
There are two limiting behaviors for $g_k$: before horizon entry 
$g_k \propto a \rightarrow h_k = \rm constant$; after horizon entry $g_k$ will oscillate which implies $h_k$ will oscillate 
with an amplitude decreasing as $1/a$.  Therefore, the current spectrum of gravitational waves is determined by the primordial 
power spectrum and by the rate at 
which scales enter the horizon (i.e. the evolution of the scale factor): $h = h_0 a_k/a$.  At horizon crossing, note that 
$k = a_k H_k$.  

During radiation domination, $H \propto a^{-2}$, 
so that $k = 1/a_k$ and during matter domination $H \propto a^{-3/2}$, so that $k = 1/a_k^{1/2}$.  From these relations, 
we find that,
\begin{eqnarray}
h_k &\propto& k^{-1} \ \rm (Radiation \; Domination) \\
h_k &\propto& k^{-2} \ \rm (Matter \; Domination)
\end{eqnarray}
In order to quantify these scalings, a transfer function is used (Turner, White \& Lidsey 1993).  
We note that during the matter dominated epoch, 
$a \propto \tau^2$ and the above equation becomes a Bessel equation (see, discussions in Pritchard \& Kamionkowski 2004).  
Requiring that the solution approach unity at $\tau =0$, we find that
\begin{equation}
h_k(\tau) = 3 \frac{j_1(k \tau)}{k \tau}
\end{equation}
Using this, 
we define the transfer function, $T(k/k_{eq})$ (where $k_{eq}$ is the wavenumber of matter-radiation equality 
$\approx 6.22 h^2 \times 10^{-2} $ Mpc$^{-1}$) as,
\begin{equation}
T(k) \equiv \sqrt{\frac{\langle h_k(k \tau)^2 \rangle}{\langle (3 j_1(k\tau)/(k \tau))^2 \rangle}}
\end{equation}
where the average is taken over several periods around the time that we want to evaluate the transfer function.  
From our scaling arguments,  the transfer function must take the form of
\begin{equation}
T(k) = \sqrt{c_0 + c_1 k/k_{eq} + c_2 (k/k_{eq})^2}
\end{equation}
The spectrum as seen by a gravitational wave detector is (Turner, White \& Lidsey 1993),
\begin{equation}
\Omega_{\mathrm{GW}}(k) = \frac{3}{128}P_T(k_0)\left( \frac{k}{k_0} \right)^{n_T}\frac{T(k/k_{eq})^2}{(k \tau_0)^2}
\end{equation}
where $\tau_0 \approx 2 H_0^{-1} \approx 10^4$ Mpc for $H_0 = 70$ km/s Mpc.  $\Omega_{\mathrm{GW}}$ is the ratio of the energy density in gravitational energy to that of the closure density, $\rho_{c} \sim 10^{-29} \ \mathrm{gm}/\mathrm{cm}^3$.  In order to determine the values for $c_1, c_2,$ and $c_3$ we need to 
numerically integrate the evolution equation 
and determine $T(k)$.  Since space-based gravitational wave detectors that are
under consideration for detection of inflationary gravitational waves have sensitivities around a
 frequency of $\sim 0.01$Hz ($\approx 10^{13}$ Mpc$^{-1}$), our analysis considers only wave numbers in excess of $k_{eq}$.  
Therefore, the values determined by Turner, White and Lidsey (1993; $c_0 = 1, c_1 = 1.34, c_2 = 2.5$) are applicable, 
even though they assumed the universe was filled with only matter and radiation.

Here, we have ignored one important detail; the right hand side of time evolution equation
 is not strictly zero.   Instead, tensor perturbations are sourced by 
anisotropic stress in the cosmic fluid.  
The largest source of such stresses comes from the free-streaming of relic neutrinos (Weinberg 2004).  
It can be shown that these neutrinos cause the amplitude of gravitational waves that enter the 
horizon well before matter radiation equality to be 
damped by a factor of $\sim 0.8$.  This will influence the amplitude of gravitational waves detectable with
space-based detectors. Similarly, the amplitude seen by CMB polarization is damped between a few to a few ten percent
with longest waves damped less.

In Fig.~4, we show the amplitude of tensor fluctuations related to a background of inflationary gravitational waves
as a function of the gravitational wave frequency. We plot roughly 25 orders of magnitude in the frequency
from super-horizon to horizon scales probed by CMB to a million-km wavelength modes detectable with
space-based interferometers. For reference, we also show various existing limits on the stochastic backgrounds
based on calculations related to the Big Bang Nucleosynthesis (Allen 1996) to timing-limit for millisecond pulsars 
(e.g., Thorsett \& Dewey 1996). An extended discussion related to such limits are in Seto \& Cooray (2005).
The left-panel of Fig.~3 shows $\Omega_{\rm GW}$ which is the fractional density contribution from a background
of stochastic gravitational waves with the density $\rho_{\rm GW}$:
\begin{equation}
\Omega_{\rm GW}(f) = \frac{1}{\rho_c} \frac{d\rho_{\rm GW}}{d \log f} \, .
\end{equation}
For reference, in the right-panel, we also plot the
dimensional strain $h_c(f)$ given by the $h_c^2(f) =   f S_h(f)$ where
$S_h(f)$ is the power spectral density of gravitational waves, $\langle h(f) h(f') \rangle = S_h(f)$.
The two quantities, $\Omega_{\rm GW}$ and $S_h$ are related through (Maggiore 2000), 
\begin{equation}
\Omega_{\rm GW}(f) = \frac{4 \pi^2}{3 H_0^2} f^3 S_h(f) \, .
\end{equation}

In addition to single detector noise levels, in Fig.~4, we also show the improvement when
 the signal from two independent detectors can be combined through a correlation analysis
between the two detectors (e.g., Maggiore 2000). Such a correlation increases 
sensitivity to a stochastic background such that,
\begin{equation}
h_{min}^{2 detector} = 1.12 \times 10^{-2}  \gamma(f)
\frac{h_n(k)}{F} \left(\frac{1 \mathrm{Hz} }{\Delta f}\right)^{1/4} \left(\frac{1  \mathrm{yr}}{T}\right)^{1/4} 
\end{equation}
where $\Delta f$ is the bandwidth over which the signals can be correlated, $T$ is the integration time,
and $F$ is a filling factor, which accounts for the fact that a primordial stochastic 
background will be isotropic on the sky, but the detector 
will only be sensitive to a fraction of the sky.  For omni-directional interferometers, such as the
Laser Interferometer Space Antenna (LISA), $F = 2/5$, while  $h_n(k)$ is the char
acteristic strain due to detector noise. The single-detector sensitivity is $h_n(k)/F$. 
Here, $\gamma(f)$ is the {\it overlap-reduction factor} that takes in to account the relative orientation
and configuration of two interferometer systems. This correction results in an overall reduction of noise
from the naive estimates that do not take in to account detail configuration and motion of the detectors with respect to each other in the orbit (e.g., Ungarelli \& Vecchio 2001).

For a correlation analysis, 
the increase in sensitivity is under the assumption that the detector noises are independent between the
two detectors, while the only correlation expected is due to stochastic signals such as inflation.
We note here that for the single  detector case, the minimum observable strain is independent of the integration time
while for a correlation analysis, long-term observations are essential.
While LISA will not allow an opportunity for such a correlation analysis, some concept studies for the
Big Bang Observer consider two systems such that the improvement related to the correlation analysis can be exploited.

In Fig.~5, we concentrate on the frequency regime relevant for space-based detectors. In addition to detector noise
and the expected level of amplitude related to inflationary gravitational waves, based on the energy scale of inflation,
we also show the confusion related to foreground stochastic backgrounds related to merging compact binaries.
These signals include cosmological neutron star-neutron star binaries (e.g., Schneider et al. 2001),
white dwarf-white dwarf binaries (e.g., Farmer \& Phinney 2003), neutron star-white dwarf binaries (e.g.,
Cooray 2004d), and the background related to single rotational neutron stars (Ferrari et al. 1999).
In addition to these compact binaries, a cosmological background of hypothetical Pop III supernovae could
potentially contribute a substantial background at frequencies corresponding to the Big Bang Observer 
(e.g., Buonanno et al. 2004).

\begin{figure}[!h]
\centerline{\psfig{file=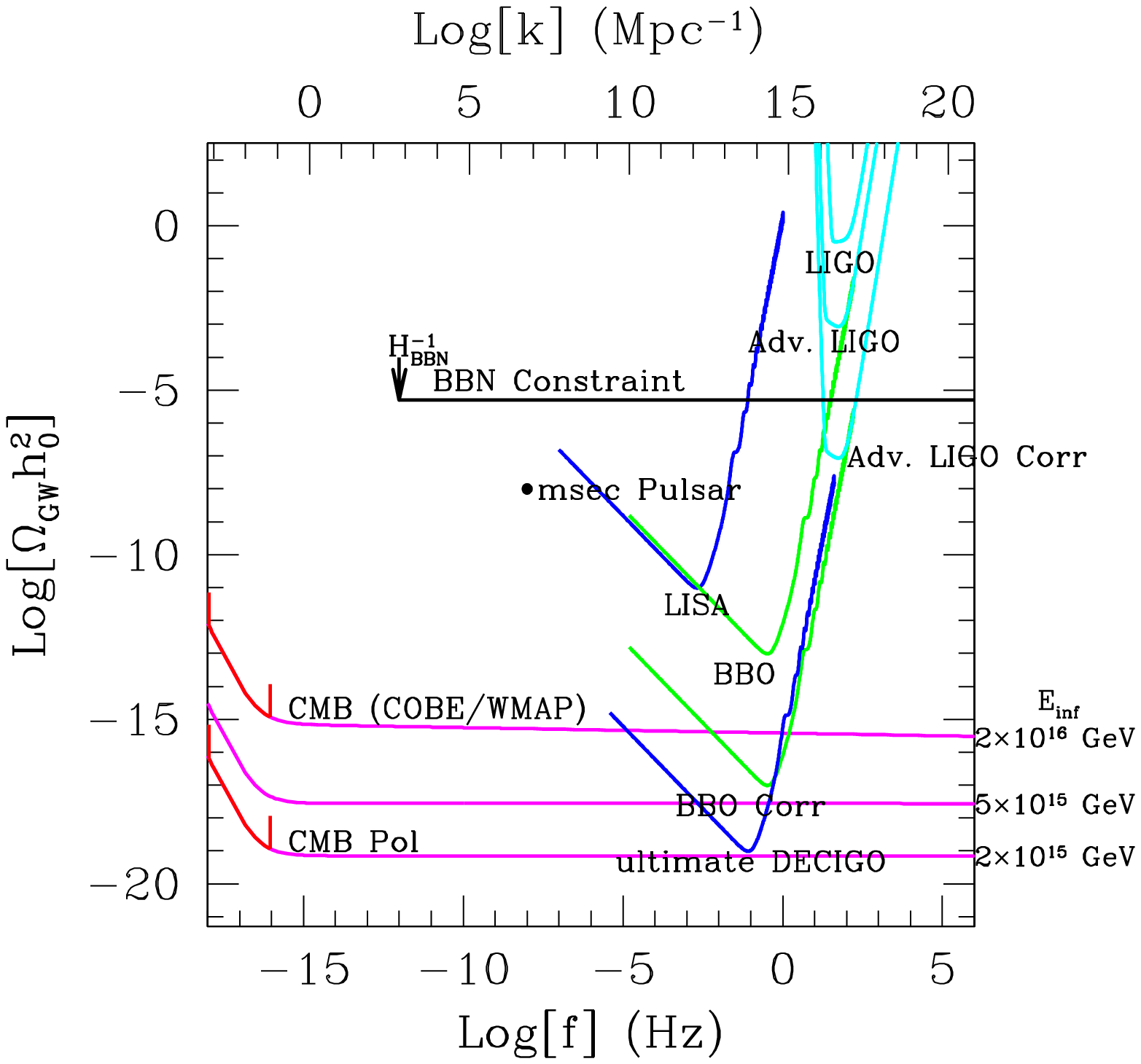,height=17pc,angle=0}
\psfig{file=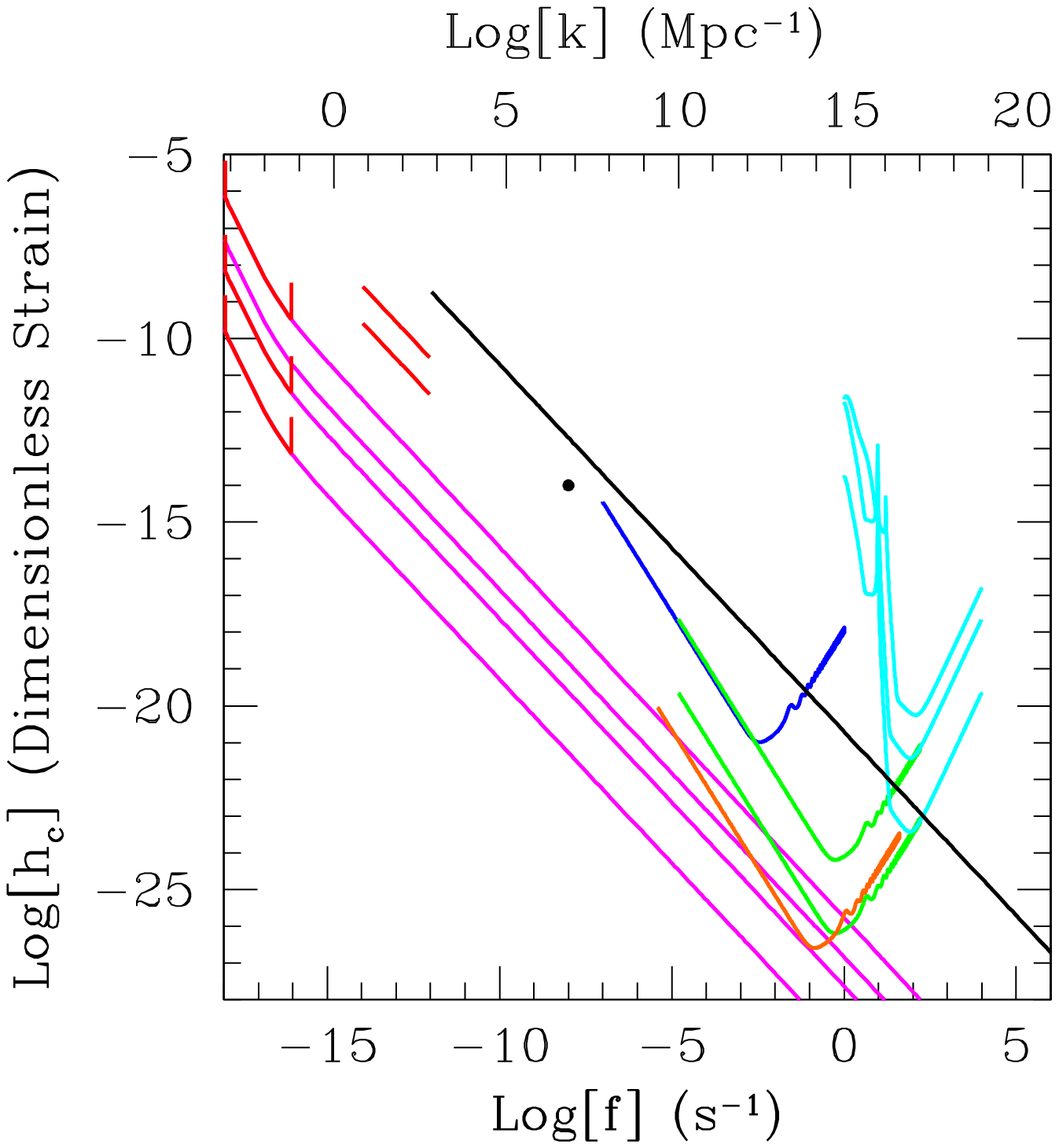,height=17pc,angle=0}}
\caption{Here we show the predicted gravitational wave background due to inflation along with various current and planned detectors and known upper limits on a cosmological stochastic gravitational wave background.  We follow Maggiore (2000) and take the correlated sensitivity to be approximately $10^{-4}$ better than the uncorrelated sensitivity.  The sensitivity of LISA is obtained from Shane Larson's website, http://www.srl.caltech.edu/$\sim$shane/sensitivity/.  The sensitivity for BBO correlated is obtained from Buonanno et al. (2004).  We estimate the sensitivity for BBO by increasing the BBO curve by 4 orders of magnitude.  The sensitivity for the ultimate DECIGO is estimated in Seto et al. (2001).  The BBN constraint comes from the requirement that the total energy density in gravitational waves be less than the energy density in other forms of radiation at BBN, so that the abundances of the light elements remain in agreement with observations (Allen 1996).  Only length scales less than the horizon size at BBO contribute and are constrained by this argument.  Millisecond pulsar timing observations have constrained the energy density in a stochastic background to be below $\sim 10^{-8}$ around $f \sim 10^{-8}$ Hz.  Curves corresponding to an inflationary produced background are shown in magenta for three different inflationary energy scales, $E_{\mathrm{inf}}$ (Turner, White \& Lidsey 1993).  As stated in the text, COBE/WMAP constrain the energy scale of inflation to be below approximately, $2 \times 10^{16}$ GeV.  As is described in the text, due to confusion with weak lensing, a detection of tensor perturbations in the polarization of the CMB can occur only if the energy scale of inflation is greater than $\sim 2 \times 10^{15}$ GeV.  As of now, there is no planned gravitational wave observatory capable of detecting a stochastic background for energies at this lower energy bound. 
}
\label{fig:joint}
\end{figure}

\begin{figure}[!h]
\centerline{\psfig{file=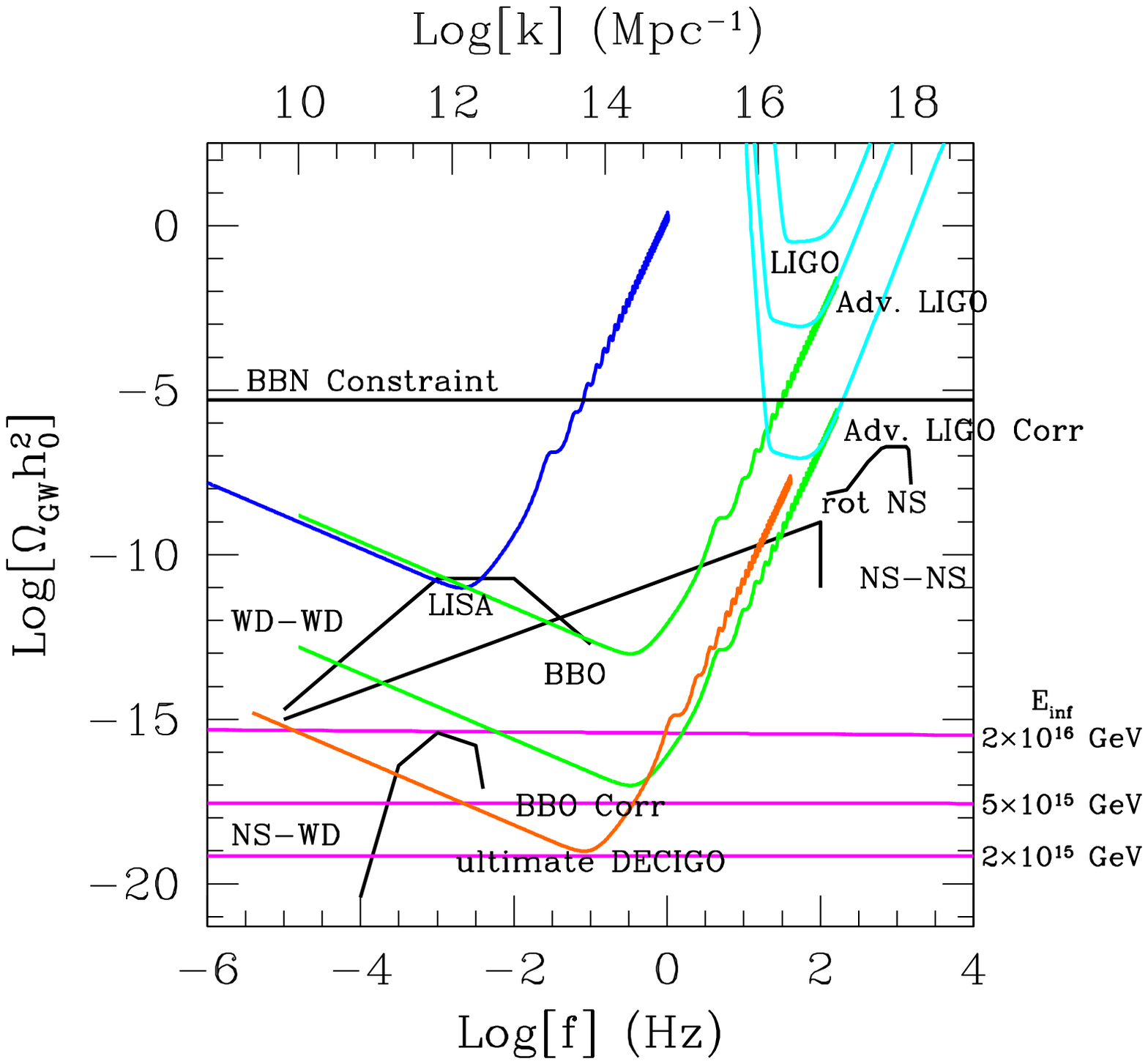,height=30pc,angle=0}}
\caption{Same as Fig.~4, but we show the background at frequencies related to the proposed Big Bang Observer mission. For comparison, we also show the expected stochastic backgrounds from foreground sources such as neutron star-neutron star binaries (e.g., Schneider et al. 2001),
white dwarf-white dwarf binaries (e.g., Farmer \& Phinney 2003), neutron star-white dwarf binaries (e.g.,
Cooray 2004d), and the background related to single rotational neutron stars (Ferrari et al. 1999).
For a reliable detection of the inflationary gravitational wave signal, the background related to neutron stars must be removed. This can be achieved given the low merger rate of these binaries and the rapid orbital 
evolution of the binary such that the gravitational wave frequency  across this band changes substantially over
the observational period.
}
\label{fig:focus}
\end{figure}

To jointly constrain the inflaton potential using observations at horizon-scale wavelengths with CMB and
million-km scale with space-based interferometers, one can extend the CMB-based technique discussed in
Turner (1993). The detectable amplitude, and the spectral index, 
with space-based detectors can be used to establish the inflaton potential amplitude and its first derivative. 
Combining this information with CMB, at roughly 30 orders of magnitude difference in scales, should
provide additional information to reconstruct the shape of the potential. 
While a priori assumed potential can easily be constrained based on the shape of the potential
al, for a general reconstruction of the potential shape, the simple approximations considered in the literature, based on power-law expansions of the potential, however, may not be appropriate given the large difference in scales. 
A detailed analysis on the extent to which inflation models can be
studies with joint constraints between CMB and direct gravitational-wave detections will be 
discussed elsewhere  (Smith et al. 2005).

{\it Acknowledgments:}
Author thanks Marc Kamionkowski, Mike Kesden, Naoki Seto, Kris Sigurdson, and Tristan Smith
for collaborative work. The author especially thanks Tristan Smith for his help with writing
review sections related to inflation and the
direct detection of inflationary gravitational waves described in figures 4 and 5. This work was
supported by the Sherman Fairchild foundation, NASA, and the Department of Energy. 


\end{document}